\documentclass[preprint,12pt]{elsarticle}
\usepackage{amssymb}

\journal{Physics Letters A}

\newcommand{\be}{\begin{equation}}
\newcommand{\ee}{\end{equation}}
\newcommand{\bea}{\begin{eqnarray}}
\newcommand{\beaa}{\begin{eqnarray*}}
\newcommand{\eea}{\end{eqnarray}}
\newcommand{\eeaa}{\end{eqnarray*}}

\begin{document}
	
\begin{frontmatter}
	
\ead{naoum@phys.uni-sofia.bg}
		
\title{\bf\Large { Antiferromagnetic to ferromagnetic phase transition as a transition to partial ordered spins. Application to $La_{1-x}Ca_xMnO_3$ in the doping range $x\geq 0.50$.}}
		
\author{N. Karchev}
		
\address{Department of Physics, Sofia University, James Bourchier 5 blvd., 1164 Sofia, Bulgaria}

\begin{abstract}
	Magnetic state is a partial ordered state	if only part of the electrons in the system give contribution to the magnetic order. 
	We study Heisenberg model of two sublattice spin system, on the body-centered cubic lattice, with antiferromagnetic nearest neighbors exchange of sublattice A and B spins 
	and two different ferromagnetic exchange constants for sublattice A ($J^A$) and B ($J^B$) spins. When $J^A>J^B$ the system undergoes transition from
	paramagnetism to ferromagnetism at Curie temperature $T_C$. Only the sublattice A spins give contribution to the magnetization of the system.
	Upon cooling, the system possesses ferromagnetism to antiferromagnetism transition at N\'eel temperature $T_N < T_C$. 
	Below  $T_N $ sublattice A and B electrons give contribution to the magnetization. The transition is a partial ordered transition. 
	There is thermodynamic evidence for this transition in the magnetic specific heat of the system. As a function of temperature there are two maxima. At high temperature $T_C$ it is $\lambda$-type. At lower temperature $T_N$ it characterizes the transition from ferromagnetism to antiferromagnetism.
	  As an example of ferromagnetism to antiferromagnetism partial ordered transition we consider the material  $La_{1-x}Ca_xMnO_3$ in the doping range $x\geq 0.50$.
     Our calculations reproduce 
     the experimental magnetization-temperature curve. 
       	
\end{abstract}


\end{frontmatter}

\section{Introduction}
\label{AFMtoFM1}

The ferromagnetic to antiferromagnetic (FM-AFM) phase transition is unusual phenomenon. 
There are various reasons causing the FM-AFM  transition. 
Experimental results and Density-functional theory calculations show an electrically manipulated FM to AFM transition in van der Waals ferromagnet  $Cr_{1.2}Te_2$ \cite{Cheng23}.
Several experiments show FM-AFM  transition driven by pressure. The contraction of the crystal under various pressures is discussed in relation to FM - AFM transition \cite{Kinoshita99}. In the single crystal of $LaCrGe_3$, increasing pressure, ferromagnetic quantum criticality is avoided by appearance of FM to AFM transition\cite{Taufour16,Taufour17}.

Perhaps the most are the experiments with chemical substitution. The FM-AFM transition in $La_{1-x}Y_xMn_2Si_2(^{57}Fe)$ compounds is realized  at $x_{cr}=0.15$. It is a result of a change in the exchange interaction linked with changes in interatomic distances \cite{Li94}.
The nature of the ferromagnetic-antiferromagnetic transition in $Y_{1-x}La_xTiO_3$ for $x< 0.3$ is reported in \cite{Hameed21}. The authors claim that  the thermal phase transition does not show conventional second-order behavior. 
Cuprate $(La-R)_4Ba_2Cu_2O_{10}$ shows FM-AFM transition by replacing $La$ ions with rare earth $(R=Nd,Sm,Eu, Gd)$ \cite{Shinya06}.

 Another material that possesses FM-AFM transition is $Pr_{1-x}Gd_xB_4$ \cite{Kobayashi09}. 
The evolution of the antiferromagnetic correlation between the magnetic moments of $Pr$ and $Gd$ breaks the ferromagnetic order of $Pr$. The experiments suggest that theoretically one has to consider two spins Heisenberg model. The temperature dependence of the magnetic specific heat of $Pr_{1-x}Gd_xB_4$  shows the main peak that corresponds to the transition from paramagnetic to ordered phase, is $\lambda$-type, indicating that the transition is of second order. The specific heat of  $Pr_{1-x}Gd_xB_4$ at (x=0.2) shows additional two peaks. They correspond to the FM-AFM transition and suggest that the transition is of first order. 

Review of the FM-AFM transition in the Heusler alloy series $Pd_2MnSn_xIn_{1-x}$ is provided in \cite{Khoi82}. Magnetic moments are localized on the $Mn$ sites. The parent compound  $Pd_2MnSn$ is a ferromagnetic with Curie temperature $T_C = 189 K $. The $Mn$ moment is $4.1\mu_B$ per atom which means that the valent state is $Mn^{3+}$. As $Sn$ is replaced by $In$ the alloy becomes multi valent and undergoes ferromagnetic to antifferomagnetic transition when  $Sn$ concentration "x" is between 0.65 and 0.45.   

The evolution of magnetism in the $UIr_{1-x}Rh_xGe$ alloy system is investigated in \cite{Haga17}. The system is ferromagnetic at $x=0.86$ and $x=0.57$. The system possesses FM-AFM transition at $x=0.55$ demonstrated by acute fall of magnetization at N\'eel temperature. The N\'eel temperature increases with decrease of $Rh$ concentration. 

We focus our attention on FM-AFM transition accompanied with transition to charge order state. The real space ordering of charge carriers in crystal occurs when Coulomb interaction overcomes the kinetic energy of carriers.
The perovskite-type manganese oxide, $Pri_{1-x}Sr_xMn0_3$  at (x = 0.5), shows  simultaneous first-order phase transition from a ferromagnetic (FM) metal to antiferromagnetic (AFM) nonmetal at 140 K  and charge ordering \cite{Tomioka95}.

The effect of the doping on the phase transition is more prominent in the experiments with $La_{1-x}Ca_xMnO_3$\cite{Schiffer95}. 
%The material is most researched for the "colossal" magnetoresistance seen at small x and low temperature in the ferromagnetic phase.
Magnetic moments are localized on the $Mn$ sites\cite{Karchev12}. The manganese has an incomplete $3d$ shell with four or three electrons for $Mn^{3+}$ and $Mn^{4+}$ respectively.
Due to the crystal field splitting effect, the degeneracy of the five $3d$ orbitals is lifted and they are grouped into one triplet $t_{2g}$ and one doublet $e_g$. The triplet has lower energy because of the space orientation of the corresponding orbitals inside 
the octahedron of six oxygen ions, surrounding the central manganese ion. 
The population of the $t_{2g}$ electrons remains constant and the Hund rule enforces alignment of the $t_{2g}$ spins into a state of maximum spin  $ s=3/2 $. 
Then the  $t_{2g}$ sector can be 
replaced by a localized spin at each manganese ion, reducing the complexity of the original five orbital model. The electrons from the $e_g$ sector however can move from ion to ion, maintaining the projection of their spin, and are called mobile electrons.    
The only important interaction between the two sectors is the Hund coupling between localized $t_{2g}$ spins and mobile $e_g$ electrons. Increasing the doping "x" the density of $e_g$ electrons decreases and charge order state emerges. 
At $x = 0.5$ the system undergoes FM-AFM transition and all $e_g$ electrons have occupied one off the antiferromagnetic
sublattices. When $x = 1$, the final compound is spin 3/2 antiferromagnetic  $CaMnO_3$.

The theoretical studies of FM-AFM transition are considerably less. Universal quantum effects are used within extended Landau theory to explain the FM-AFM transition\cite{Belitz17}. On the other hand, it is shown that microscopic properties of specific materials trigger such transition. Effective models incorporating these properties are suitable tools to study magnetic properties of the materials.  Low energy model of $USb_2$ is derived to study the FM-AFM transition \cite{Marcin18}. Novel mechanism of FM-AFM transition is proposed\cite{Marcin19} to explain the experimentally observed pressure induced transition in $LaCrGe_3$ as a mechanism to avoid the ferromagnetic quantum criticality\cite{Taufour16,Taufour17}.

In the present article we consider FM-AFM transition as a partial ordered transition. Magnetic state is a partial ordered state if only part of the electrons in the system give contribution to the magnetic order. It is studied in exactly solvable models \cite{Vaks66,Azaria87,Diep04}, by means of Green's function approach \cite{Diep97} and utilizing the Monte Carlo method \cite{Azaria87}.
A modified spin-wave theory of magnetism has been developed to investigate the state of partial order in spin-fermion systems \cite{Karchev08} and field-cooled ferrimagnetic spinels \cite{Karchev15}. In the present paper, we use this theory to study the transition from antiferromagnetism to ferromagnetism in $La_{1-x}Ca_xMnO_3$ in the doping range $x\geq 0.50$. It is proved that this transition is partial order one. The phase diagram of $La_{1-x}Ca_xMnO_3$ in the doping range $x<0.50$ is theoretically studied in \cite{Karchev12}.

\section{Model}
\label{AFMtoFM2}

We consider Heisenberg model of two sublattice spin system with antiferromagnetic nearest neighbors exchange $(J)$ of sublattice A and B spins 
%$({\bf S}^A,{\bf S}^B)$
 and two different ferromagnetic exchange constants for sublattice A $(J^A)$ and B $(J^B)$ spins. The Hamiltonian of the system is 
\bea \label{FMAFM1}
H & = & J \sum\limits_{\langle ij \rangle} {{\bf S}^A_{i}
	\cdot {\bf S}^B_{j}} \\
 & & - J^A\sum\limits_{\ll ij \gg _A } {{\bf S}^A_{i}
	\cdot {\bf S}^A_{j}}\,-\,J^B\sum\limits_{\ll ij \gg _B } {{\bf S}^B_{i}
	\cdot {\bf S}^B_{j}}\nonumber
\eea 
where $({\bf S}^A,{\bf S}^B)$ are spin $s=3/2$ operators, the sums are over all sites of a body-centered cubic (bcc) lattice. To be specific we choose $(J^A>J^B)$. We introduce Holstein-Primakoff representation for the spin
operators ${\bf S}^A_{i}(a^+,a)$ and ${\bf S}^B_{i}(b^+,b)$. Rewriting the effective Hamiltonian in terms of the Bose operators $(a^+,a,b^+,b)$ we keep  only the quadratic and quartic terms. The next step is to represent the Hamiltonian in the Hartree-Fock (HF)  approximation, with temperature-dependent HF parameters to be determined self consistently: 
%\cite{supp}. 
\be
H \approx H_{HF}=H_{cl}+H_q, \label {FMAFM7} \ee
where
\bea\label{FMAFM8} H_{cl} & = & 6 N J^A s^2 (u^A-1)^2+ 6 N J^B s^2 (u^B-1)^2 \nonumber \\
& + & 8 N J s^2 (u-1)^2,
\eea
and 
\begin{equation}\label{FMAFM9}
H_q = \sum\limits_{k\in B_r}\left [\varepsilon^a_k\,a_k^+a_k\,+\,\varepsilon^b_k\,b_k^+b_k\,-
\,\gamma_k \left (a_k^+b_k^+ + b_k a_k \right )\,\right ],
\end{equation}
with $N=N^A=N^B$ is the number of sites on a sublattice. The two equivalent sublattices A and B of the bcc lattice are simple cubic lattices. The wave vector $k$ runs over the reduced  first Brillouin zone $B_r$ of a bcc lattice which is the first Brillouin zone of a simple cubic lattice. 

The dispersions are given by equalities
\begin{eqnarray}\label{FMAFM10}
\varepsilon^a_k & = & 4s J^A u^A \left(3-\cos k_x-\cos k_y - \cos k_z\right)
\,+\,8s\,J u \nonumber\\
\varepsilon^b_k & = & 4s J^B u^B \left(3-\cos k_x-\cos k_y - \cos k_z\right)
\,+\,8s\,J u \nonumber\\
\gamma_k & = & 8J\,u\,s\,\cos \frac {k_x}{2} \,\cos \frac {k_y}{2} \,\cos \frac {k_z}{2} \end{eqnarray}
The equations (\ref{FMAFM10}) show that Hartree-Fock parameters ($u^A,u^B,u$) renormalize the intra and inter-sublattice exchange constants
($J^A, J^B,J$) respectively.

To diagonalize the Hamiltonian one introduces new Bose fields
$\alpha_k,\,\alpha_k^+,\,\beta_k,\,\beta_k^+$ by means of Bogoliubov transformation for Bose system:
\begin{eqnarray}\label{fc-ferri6} & &
a_k\,=u_k\,\alpha_k\,+\,v_k\,\beta^+_k\qquad
a_k^+\,=u_k\,\alpha_k^+\,+\,v_k\,\beta_k
\nonumber \\
\\
& & b_k\,=\,u_k\,\beta_k\,+\,v_k\,\alpha^+_k\qquad
b_k^+\,=\,u_k\,\beta_k^+\,+\,v_k\,\alpha_k, \nonumber
\end{eqnarray}
where the coefficients of the transformation $u_k$ and $v_k$ are real functions of the wave vector $k$:
\bea \label{ferri11} &
&u_k\,=\,\sqrt{\frac 12\,\left (\frac
	{\varepsilon^a_k+\varepsilon^b_k}{\sqrt{(\varepsilon^a_k+\varepsilon^b_k)^2-4\gamma^2_k}}\,+\,1\right
	)}\nonumber \\
\\
& & v_k\,=\,sign (\gamma_k)\,\sqrt{\frac 12\,\left (\frac
	{\varepsilon^a_k+\varepsilon^b_k}{\sqrt{(\varepsilon^a_k+\varepsilon^b_k)^2-4\gamma^2_k}}\,-\,1\right
	)}.\nonumber \eea

 The transformed Hamiltonian adopts the form
\begin{equation}
\label{FMAFM2} H_q = \sum\limits_{k\in B_r}\left
(E^{\alpha}_k\,\alpha_k^+\alpha_k\,+\,E^{\beta}_k\,\beta_k^+\beta_k\,+\,E^0_k\right),
\end{equation}
with new dispersions
\begin{eqnarray}  \label{fc-ferri8}
 & & E^{\alpha}_k\,=\,\frac 12\,
\left[
\sqrt{(\varepsilon^a_k\,+\,\varepsilon^b_k)^2\,-\,4\gamma^2_k}\,-\,\varepsilon^b_k\,+\,\varepsilon^a_k\right] \nonumber \\
\\
& & E^{\beta}_k\,=\,\frac
12\,\left [
\sqrt{(\varepsilon^a_k\,+\,\varepsilon^b_k)^2\,-\,4\gamma^2_k}\,+\,\varepsilon^b_k\,-\,\varepsilon^a_k\right]
\nonumber \end{eqnarray}
and vacuum energy
\begin{equation}\label{fc-ferri9}
E^{0}_k\,=\,\frac
12\,\left [
\sqrt{(\varepsilon^a_k\,+\,\varepsilon^b_k)^2\,-\,4\gamma^2_k}\,-\,\varepsilon^b_k\,-\,\varepsilon^a_k\right]
\end{equation}
 The wave vector $k$ runs over the reduced  first Brillouin zone $B_r$ of a bcc lattice .

For positive values of the Hartree-Fock parameters and all values of $k\in B_r$,
the dispersions are nonnegative $ E^{\alpha}_k\geq 0,\, E^{\beta}_k \geq 0$.
The $\alpha_k$ and $\beta_k$ bosons are the long-range \textbf{(magnon)} excitations in the system with antiferromagnetic dispersion $E^{\alpha}_k\propto c^{\alpha} |k|$ and $E^{\beta}_k\propto c^{\beta} |k|$, near the zero wave vector\cite{supp}. For spin velocity constants we obtained
\begin{equation}\label{FMAFM3}
c^{\alpha} = c^{\beta} = 4s\sqrt{Ju\left(J^A u^A + J^B u^B + J u\right)}.
\end{equation}

The free energy of a system with Hamiltonian $H_{HF}$ equations (\ref{FMAFM7}), (\ref{FMAFM8}) and  (\ref{FMAFM9}) is
\begin{eqnarray}\label{fc-ferri10}
\mathcal{F} & = & 6 N J^A s^2 (u^A-1)^2+ 6 N J^B s^2 (u^B-1)^2 \nonumber \\
& + & 8 N J s^2 (u-1)^2 + \frac 1N \sum\limits_{k\in B_r}E^{0}_k \\
& + & \frac {1}{\beta N} \sum\limits_{k\in B_r}\left[ \ln\left(1-e^{-\beta E^{\alpha}_k}\right)\,+\,\ln\left(1-e^{-\beta E^{\beta}_k}\right)\right],\nonumber\end{eqnarray}
where $\beta\,=\,1/T$\,\, is the inverse temperature, with Boltzmann constant $k_B$ set equal to 1 . 
Then, the system of equations for the Hartree-Fock parameters is
\begin{equation}\label{fc-ferri11}\partial\mathcal{F}/\partial u^A=0,\quad \partial\mathcal{F}/\partial u^B=0,\quad\partial\mathcal{F}/\partial u=0.\end{equation}
One can write them by means of the Bose functions $n_k^{\alpha}$ and $n_k^{\beta}$ of $\alpha$ and $\beta$ excitations \cite{supp}.

The Hartree-Fock parameters are positive functions of $T/J$ , solution of the system of equations (\ref{fc-ferri11}). Utilizing these functions, one can calculate the spontaneous magnetization on the two sublattices
\begin{eqnarray} \label{fc-ferri12}
M^A & = & <S^{A3}_{j}> \,\,\, j\,\, is\,\, from\,\, sublattice\,\, A \nonumber \\
\\
M^B & = & <S^{B3}_{j}> \,\,\, j\,\, is\,\, from\,\, sublattice\,\, B \nonumber \end{eqnarray}
and $M\,=\,M^A\,+\,M^B$, the spontaneous magnetization of the system.
In terms of the Bose functions 
\begin{eqnarray}\label{fc-ferri13}
M^A & = & s\,-\,\frac 1N \sum\limits_{k\in B_r} \left[u_k^2 \,n_k^{\alpha}\, +\, v_k^2\, n_k^{\beta}\, +\, v_k^2\right] \\
M^B & = & - \,s\,+\,\frac 1N \sum\limits_{k\in B_r} \left[v_k^2 \,n_k^{\alpha}\, +\, u_k^2\, n_k^{\beta}\, +\, v_k^2\right],\nonumber \end{eqnarray}
where $u_k$ and $v_k$ are functions of the wavevector $k$, Bogoliubov coefficients in the transformation \cite{supp}.

We utilize the solutions of HP parameters as a functions of temperature to calculate the sublattice A and B magnetizations. The magnon excitations - $\alpha_k$ and $\beta_k$ are complicated mixtures of the transversal
fluctuations of the $A$ and $B$ spins. As a result the magnons' fluctuations suppress in a different way the magnetization on sublattices $A$ and $B$. Quantitatively this depends on the exchange constants $J^A$ and $J^B$ . At characteristic temperature $T_N$  spontaneous magnetization on sublattice $B$ becomes equal to zero, while spontaneous magnetization on sublattice $A$ is still nonzero\cite{supp}. The total magnetization $M^A+M^B$ increases and reaches its maximum value at $T_N$. The system in the temperature interval $(0,T_N)$ has two antiferromagnetic magnons but sublattice A and B magnetizations do not compensate each other. We call this phase noncompensated antiferromagnetic.     

Above the N\'eel temperature the magnetic moments of the sublattice B electrons do not contribute the magnetization of the system. Only sublattice A electrons do. To study this "partial order" we make use of the Takahashi modified spin-wave theory \cite{Takahashi87,Karchev08,Karchev15} and introduce two parameters $\lambda^A$ and $\lambda^B$  to enforce the magnetization on the two sublattices to be equal to zero in paramagnetic phase. The new Hamiltonian is obtained from the original  one by adding two new terms:
\begin{equation}
\label{fc-ferri14} \hat{H}\,=\,H\,-\,\sum\limits_{i\in A}
\lambda^A S^{A3}_{i}\,+\,\sum\limits_{i\in B} \lambda^B S^{B3}_{i} \end{equation}   
In momentum space
the new Hamiltonian adopts the form \be \label{ferri27}
\hat{H} = \sum\limits_{k\in B_r}\left [\hat{\varepsilon}^a_k\,a_k^+a_k\,+\,\hat{\varepsilon}^b_k\,b_k^+b_k\,-
\,\gamma_k\,(b_k a_k+b_k^+ a_k^+)\right] \ee
where the new dispersions are
\be \label{ferri28}
\hat{\varepsilon}^a_k\,=\varepsilon^a_k\,+\,\lambda^A, \qquad
\hat{\varepsilon}^b_k\,=\varepsilon^b_k\,+\,\lambda^B.\ee
It is convenient to represent the parameters
$\lambda^A$ and $\lambda^B$ in the form
\begin{equation} \label{fc-ferri15}
\lambda^A\,=\,8 J u s (\mu^A\,-\,1),\quad
\lambda^B\,=\,8 J u s (\mu^B\,-\,1). \end{equation}
The dispersions $\hat{\varepsilon}^a_k$ and $\hat{\varepsilon}^b_k$ adopt the form 
\begin{eqnarray}\label{FMAFM20}
\hat{\varepsilon}^a_k & = & 4s J^A u^A \left(3-\cos k_x-\cos k_y - \cos k_z\right)
\,+\,8s\,J u \mu^A \nonumber\\
\\
\hat{\varepsilon}^b_k & = & 4s J^B u^B \left(3-\cos k_x-\cos k_y - \cos k_z\right) 
\,+\,8s\,J u \mu^B. \nonumber  
\end{eqnarray}
They are positive ($\hat{\varepsilon}^a_k>0$, $\hat{\varepsilon}^b_k>0$) for all values of the wavevector $k$, if the parameters  $\mu^A$ and $\mu^B$ are positive ($\mu^A>0,\,\mu^B>0$).

To obtain the Hamiltonian (\ref{ferri27}) in diagonal form, we use the Bogoliubov transformation 
\cite{supp} 
replacing $\varepsilon^a_k$ and  $\varepsilon^b_k$ with  $\hat{\varepsilon}^a_k$ and $\hat{\varepsilon}^b_k$. The result is
\be
\label{ferri30} \hat{H} = \sum\limits_{k\in B_r}\left
(\hat{E}^{\alpha}_k\,\alpha_k^+\alpha_k\,+\,\hat{E}^{\beta}_k\,\beta_k^+\beta_k+\hat{E}^0_k\right),
\ee 
where
\bea\label{ferri31} & & \hat{E}^{\alpha}_k\,=\,\frac
12\,\left [
\sqrt{(\hat{\varepsilon}^a_k\,+\,\hat{\varepsilon}^b_k)^2\,-\,4\gamma^2_k}\,-\,\hat{\varepsilon}^b_k\,+\,\hat{\varepsilon}^a_k\right] \nonumber \\
& & \hat{E}^{\beta}_k\,=\,\frac
12\,\left [
\sqrt{(\hat{\varepsilon}^a_k\,+\,\hat{\varepsilon}^b_k)^2\,-\,4\gamma^2_k}\,+\,\hat{\varepsilon}^b_k\,-\,\hat{\varepsilon}^a_k\right]\\
& & \hat{E}^{0}_k\,=\,\frac
12\,\left [
\sqrt{(\hat{\varepsilon}^a_k\,+\,\hat{\varepsilon}^b_k)^2\,-\,4\gamma^2_k}\,-\,\hat{\varepsilon}^b_k\,-\,\hat{\varepsilon}^a_k\right]\nonumber\eea
The dispersions Eq.(\ref{ferri31}) are well defined if square-roots in equations
(\ref{ferri31}) are well defined. This is true if 
\be\label{ferri34} \mu^A\mu^B\geq1.\ee
If $\mu^A\mu^B >1$ the dispersions (\ref{ferri31}) are positive ($\hat{E}^{\alpha}_k>0, \hat{E}^{\beta}_k>0$). This means that the system is in paramagnetic phase. 
When  $\mu^A\mu^B =1$ the spectrum of the system posseses long-range (magnon) excitation and it is in ordered phase. 

In the partial ordered phase above $T_N$ with $M^B=0$ sublattice A magnetization contributes the total magnetic moment. To study this ordered phase one has to solve a system of four equations, three for Hartree-Fock paraneters and one $(M^B=0)$ for $\mu^A=1/\mu^B$. The solution of the system, the Hartree-Fock parameters and parameters $\mu^A=1/\mu^B$ as fuctions of $T/J$ within interval $(0,T_C/J)$ are depicted in \cite{supp}. Above the $T_C$, the critical order-disorder transition temperature, one obtains $\mu^A\mu^B>1$. The sublattice A and B magnetization $M^A$ and $M^B$ in partial ordered phase  are depicted in \cite{supp}. The total magnetization $M^A+M^B$ is depicted in Fig.(1).

The figure for  $\mu^A$ and $\mu^B$ \cite{supp} shows that $\mu^B$ is larger than $\mu^A$. As a result we obtain that  $\beta$ excitation is gapped ($E^{\beta}_k>0$ for all values of the wave vector $k$), while $E^{\alpha}_0=0$ and near the zero wave vector
\be \label{ferri35b}
\hat{E}^{\alpha}_k\propto \hat{\rho} k^2\ee
with spin-stiffness constant
\bea \label{ferri35c} \hat{\rho} & = & s\frac {(\mu^A+\mu^B)(J^A u^A +J^B u^B)+2JU}{ \mu^B-\mu^A}  \nonumber \\
& + & sJ^A u^A -sJ^B u^B.\eea
This means that  $\alpha_k$ boson is the long-range excitation (ferromagnetic magnon) in the system.

\begin{figure}[!ht]
	\centering\includegraphics[width=4in]{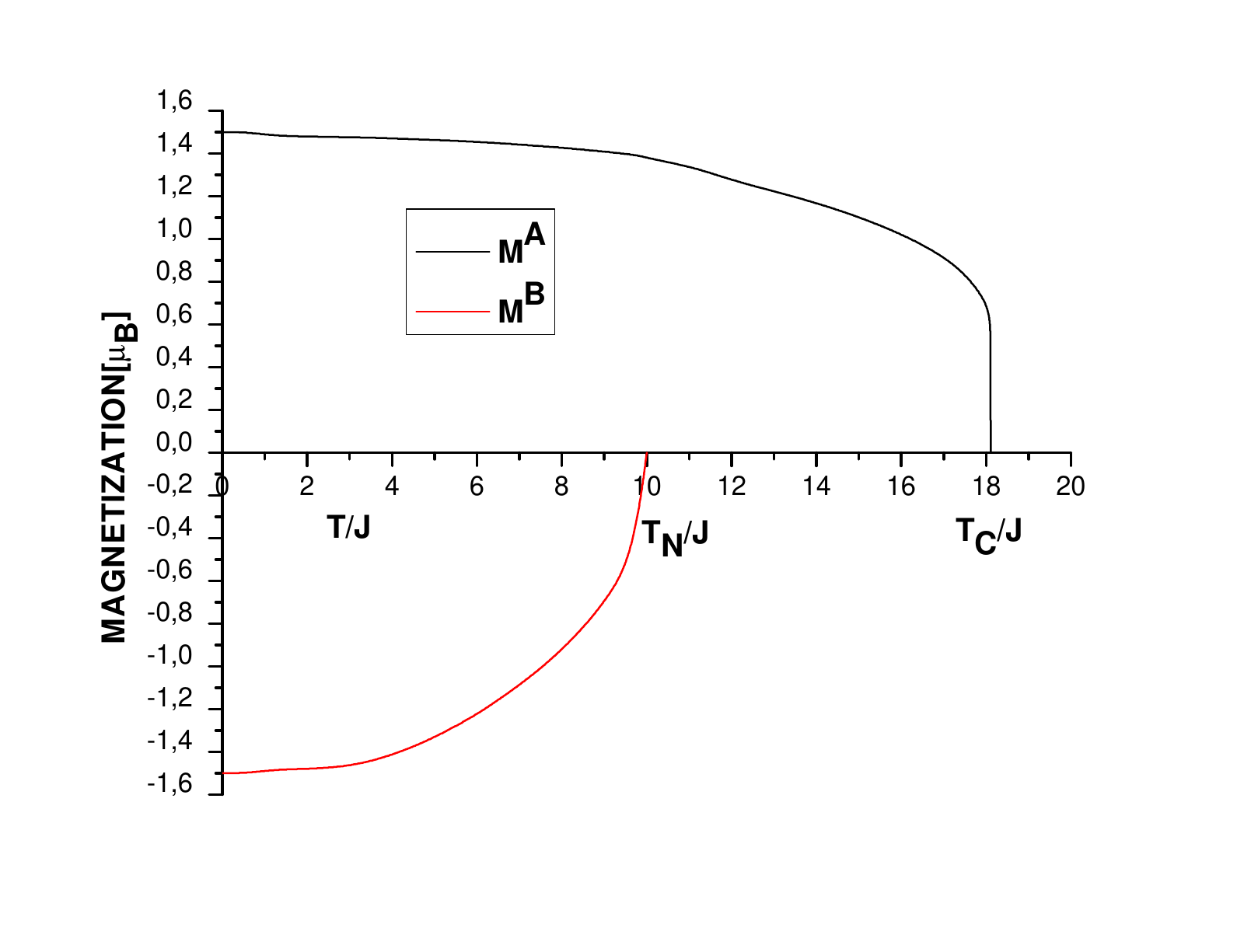}
	%	\epsfxsize=\linewidth
	%	\epsfbox{(fig5)SuppAFMtoFM-MM}
	%	\vskip -6 cm
	\caption{The sublattice $A$ and $B$ magnetization are depicted as a function of dimensionless temperature $T/J$ within the interval $(0,T_C/J)$, for a system with parameters $J^A/J=0.8$ and  $J^B/J=0.006$.}\label{fig-MAMB}       
\end{figure}

The sublattice $A$ and $B$ magnetizations are depicted as a function of dimensionless temperature $T/J$ within the interval $(0,T_C/J)$, for a system with parameters $J^A/J=0.8$ and  $J^B/J=0.006$ in figure (\ref{fig-MAMB}). The figure shows that at $T_N$ the system undergoes partial order transition from low temperature antiferromagnetic phase with $M^A>0$ and $M^B<0$, to high temperature ferromagnetic phase with $M^A>0$ and $M^B=0$.   

By combining the calculations for the two phases, we obtain the total magnetization of the system as a function of temperature. 
As the temperature decreases, the onset of magnetism is at $T_C$. Below $T_C$ the magnetization increases and reaches the maximum value at $T_N$. Below $T_N$ magnetization decreases abruptly or smoothly depending on the parameters in the model and approaches zero. In the high temperature phase$ (T_N,T_C)$ the system has one magnon with dispersion proportional to $k^2$  and the total magnetization is equal to the magnetization of sublattice A. The partial ordered phase is ferromagnetic. In the low temperature phase $(0,T_N)$ the system has two magnons with dispersion proportional to $|k|$  and the total magnetization $M=M^A+M^B$  is nonzero. The phase is noncompencated antiferromagnetic.

\begin{figure}[tb]
	\begin{center}
		\includegraphics[width=100mm]{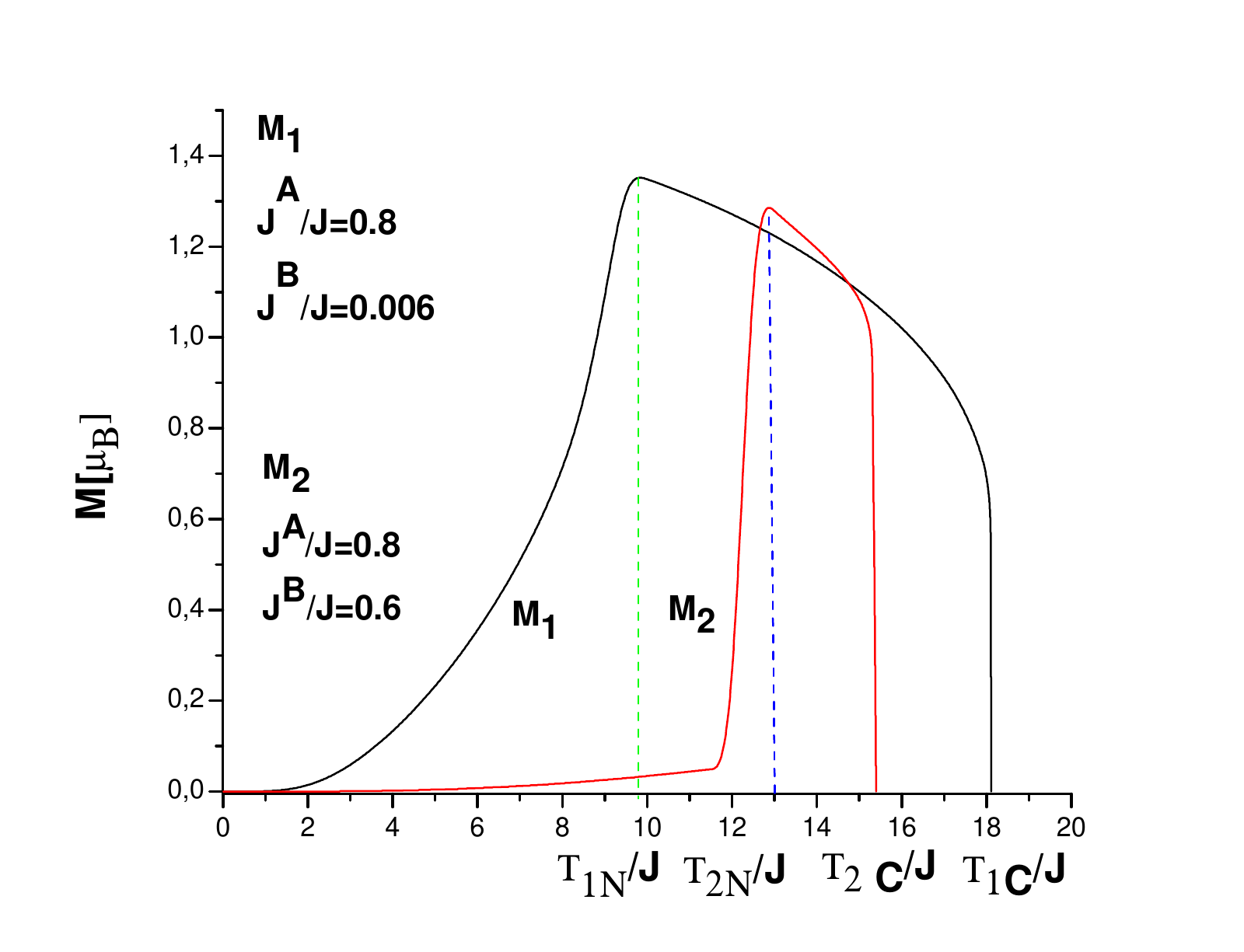}
	\end{center}
\caption{The magnetization $M^A+M^B$ are depicted, in units of $\mu_B$, as a function of dimensionless temperature $T/J$ for systems $(M_1)$ with parameters $J^A/J=0.8, J^B/J=0.006$ and $(M_2)$ with parameters  $J^A/J=0.8, J^B/J=0.6$. The $(M_1)$ system has a partial ordered transition from ferromagnetic to noncompensated antiferromagnetic phase at  $T_{1N}/J$ (green dash line).  The $(M_2)$ system has a partial ordered transition from ferromagnetic to antiferromagnetic phase at  $T_{2N}/J$ (blue dash line). }\label{fig1-MM}       
\end{figure}

The curves for systems with parameters $(J^A/J=0.8,J^B/J=0.006)$ and  $(J^A/J=0.8,J^B/J=0.6)$ are plotted in figure (\ref{fig1-MM}). They show that upon cooling the system has a transition from paramagnetic to ferromagnetic phase at $T_{1C}$ and $T_{2C}$ and at lower temperature $T_{1N}$ and $T_{2N}$ a transition from ferromagnetic to antiferromagnetic phase.

For the system $(M_1)$ with a sublattice A exchange constant $J^A$ much larger than $J^B$, the ferromagnetic "partial ordered" phase  dominates.  It is "partial ordered"  phase because only sublattice A electrons contribute the magnetization of the system. Below $T_{1N}$ the magnetization smoothly decreases and approaches zero. This phase is noncompensated antiferromagnetic and the system undergoes transition from ferromagnetic to noncompensated antiferromagnetic phase which is partial ordered transition.  

For  the system  $(M_2)$ the values of the exchange constants $J^A$ and $J^B$ are very close. The low temperature phase $(0,T_{2N})$ dominates. All electrons of the system contribute the total magnetization which is zero. This means that the phase is antiferromagnetic, and at $T_{2N}$ the system undergoes transition from antiferromagnetic to ferromagnetic phase. It is partial ordered transition.

\begin{figure}[tb]
\begin{center}
\includegraphics[width=100mm]{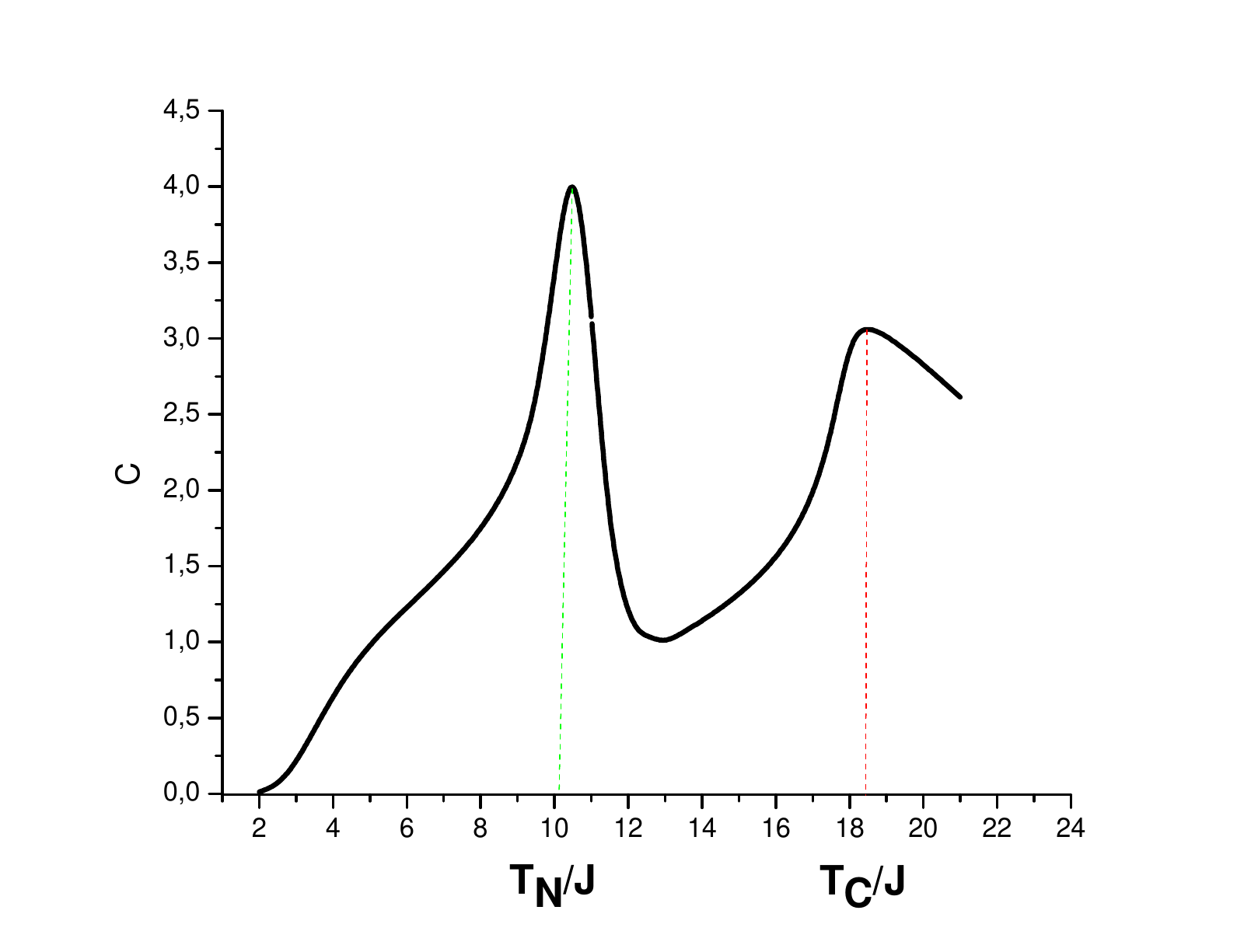}
\end{center}
%\begin{figure}[!ht]
%\centering\includegraphics[width=3.7in]{(fig2)C}
%\epsfxsize=\linewidth
%\epsfbox{_fig2_C}
%\vskip -2.5 cm
\caption{ The magnetic specific heat is depicted as a function of dimensionless temperature $T/J$ for a system with parameters $J^A/J=0.8$ and $J^B/J=0.006$.The high-temperature peak at $T_C/J$ is a consequence of paramagnetic to ferromagnetic phase transition. It is $\lambda$-type second order transition. The peak at   $T_N/J$ characterizes the partial ordered ferromagnetic to noncompensated antiferromagnetic phase transition. The small shoulder below $T_N$ is a sign for a first order transition }\label{fig2-C}
\end{figure}

An important feature of the FM-AFM transition is the magnetic specific heat of the system. The specific heat $C(T/J)$ as a function of dimensionless temperature $T/J$ is depicted in figure (\ref{fig2-C}). 
The function has two peaks. At $T_C/J$ characterizes paramagnetic to ferromagnetic phase transition and at $T_N/J$ the ferromagnetic to antiferromagnetic phase transition. At $T_C/J$, the order disorder transition is $\lambda$ -type, suggesting that it is second order. The small shoulder below  $T_N/J$  indicates that it is of first order. In the compound $Pr_{1-x}Gd_xB_4$ the first-order transition is more expressively  demonstrated by two peaks\cite{Kobayashi09}.

We apply the above developed theory of FM-AFM transition accompanied with transition to charge ordered state for $La_{1-x}Ca_xMnO_3$ in the doping range $x\geq 0.50$. At x=0.5, all the manganese ions that occupy the sites of the A sublattice are in the $Mn^{3+}$ state, while those that occupy the sites in the B sublattice are in the  $Mn^{4+}$ state. The charge ordered  state is in the heart of the FM-AFM transition. When x=1, i.e the density of charged electrons is zero, the system is spin $s=3/2$ antiferromagnetic parent compound $CaMnO_3$. One can describe it by spin 3/2 spin operators ${\bf S}^A_i$ , ${\bf S}^B_j$ and Heisenberg Hamiltonian of an antiferromagnetic system. 

Within the interval $0.5<x<1$, all $e_g$ electrons occupy the sites of the sublattice A, and by their averaging we obtain two additional spin-spin ferromagnetic exchange terms in the Heisenberg Hamiltonian with $J^A$ exchange constant greater than $J^B$. As $x$ increases, the density of $e_g$ electrons decreases, therefore the effective exchange constants $J^A$ and $J^B$ decrease preserving the inequality $J^A>J^B$.

\begin{figure}[tb]
\begin{center}
\includegraphics[width=100mm]{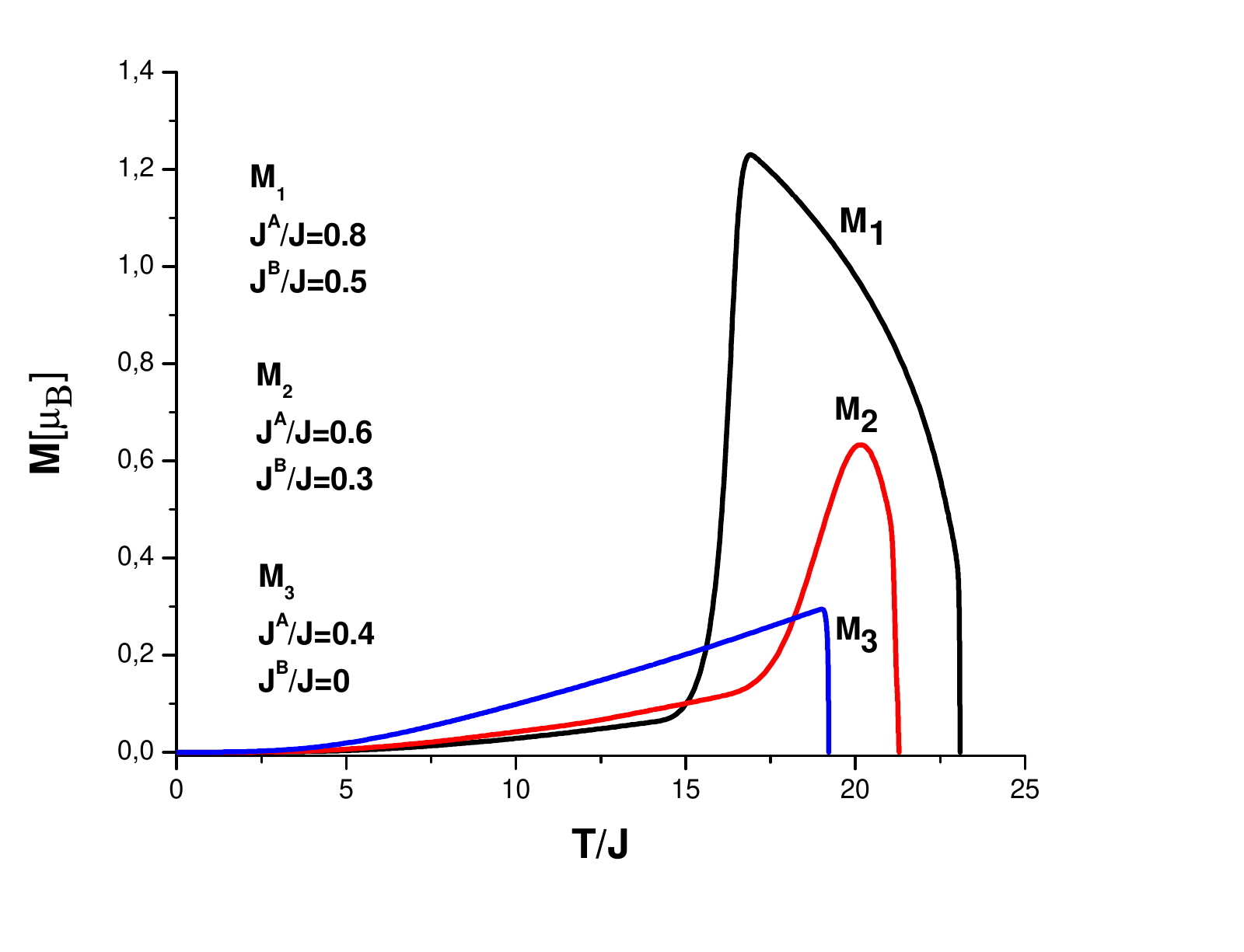}
\end{center}
%\begin{figure}[!ht]
%\centering\includegraphics[width=4.7in]{(fig3)exp}
%\epsfxsize=\linewidth
%\epsfbox{_fig3_exp.pdf}
%\vskip -2.5 cm
\caption{The magnetization-temperature curves are depicted in figure for systems:  $M_1$ with parameters $J^A/J=0.8$ and $J^B/J=0.5$,  $M_2$ with parameters $J^A/J=0.6$ and $J^B/J=0.3$ and $M_3$ with parameters $J^A/J=0.4$ and $J^B/J=0$. Compare with the experimentally measured magnetization temperature curves of $La_{1-x}Ca_xMnO_3$ \cite{Schiffer95} one obtains that $M_1$ curve  reproduces to grate extend the experimental magnetization-temperature curve for  $x=0.50$, $M_2$ is in good agreement when $x=0.52$ and $M_3$ qualitatively matches the experimental curve at $x=0.55$. }\label{fig3-exp}
\end{figure}

We explore three systems, $M_1$ with parameters $J^A/J=0.8$ and $J^B/J=0.5$, $M_2$ with parameters $J^A/J=0.6$ and $J^B/J=0.3$ and $M_3$ with parameters $J^A/J=0.4$ and $J^B/J=0$. The magnetization-temperature curves are depicted in figure (\ref{fig3-exp}).  

By comparison with the experimentally measured magnetization temperature curves of $La_{1-x}Ca_xMnO_3$ \cite{Schiffer95} depicted in Fig.4, we find that our calculations for system $M_1$ reproduce to grate extend the experimental magnetization-temperature curve for  $x=0.50$, those for $M_2$ are in good agreement when $x=0.52$ and system $M_3$ qualitatively match the experimental curve at $x=0.55$. The differences are in the tails when the temperature reaches the critical value. They are consequence of the weak magnetic field $H=4T$ applied during the measurement which makes the slope of the curve M(T) flater. It is not accounted for in the calculations.

The $M_1$ curve coincides with the experimental one of $Pr_{1-x}Sr_xMnO_3$ \cite{Tomioka95} for $x=0.5$.
Finally, all three curves reproduce the results for $Pr_{1-x}Gd_xB_4$ \cite{Kobayashi09} when $x=0.2$, $x=0.22$ and $x=0.25$. This suggests that applying magnetic field during preparation of $La_{1-x}Ca_xMnO_3$ one can destroy the AFM-FM transition. 

\section{Conclusion}
\label{AFMtoFM3}

In conclusion, we have studied the antiferromagnetic to ferromagnetic phase transition as a transition to partial ordering of spins triggered by charge ordering. As example we have considered multivalent manganites $La_{1-x}Ca_xMnO_3$. When $x<0.5$ the system undergoes paramagnetic to ferromagnetic phase transition \cite{Schiffer95}. The phase diagram of $La_{1-x}Ca_xMnO_3$ in the doping range $x<0.50$ is theoretically studied in \cite{Karchev12}. Charge ordered state emerges in the doping range  $x\geq 0.50$. It introduces two sublattices A and B and  all sites of A are occupied by manganies ions in $Mn^{3+}$ state, so that 
manganies $e_g$ mobile electrons occupy sublattice A. At $x=0.5$ the system undergoes transition to charge ordered state accompanied with FM-AFM transition. Experiments \cite{Schiffer95} show that decreasing the temperature the system undergoes paramagnetic to ferromagnetic phase transition. Our calculations have supplemented  this result with the conclusion that only sublattice A spins contribute the magnetisation thereby the ferromagnetic phase is partial ordered state. The magnetization increases with decreasing temperature, reaching its maximum value at $T_N$. Below N\'eel temperature it abruptly falls to zero. Our result have added that in this phase the system has two magnons with antiferromagnetic linear dispersion and all spins of the system contribute magnetization. Hence, the ferromagnetism to antifferomagnetism phase transition is partial ordered transition.   

When the doping x is increased, the experiment shows that the shape of magnetisation temperature curve changes. Upon cooling the magnetisation increases reaching the maximum at $T_N$. Below this temperatire the magnetisation decreases slowly approaching zero at low temperature. This means that the total magnetization is not zero. 
Our results have demonstrated that the system has two antiferromagnetic magnons, hence the system is in noncompensated antiferromagnetic phase.

The theoretical studies have reproduced very well the nontrivial fact that the transition from ferromagnetism to antiferromagnetism changes into a transition from ferromagnetism to uncompensated antiferromagnetism when $x$ increases.

The experimental magnetization-temperature curves for $Pr_{1-x}Gd_xB_4$ \cite{Kobayashi09} and $Pr_{1-x}Sr_xMnO_3$ \cite{Tomioka95} match very well the theoretical reported in the present paper.

The partial order transition has been observed in $Ca_3CoRhO_6$ \cite{Kageyama98,Niitaka99} compound which contains one-dimensional chains with ferromagnetic exchange in the chain and antiferromagnetic between chains . The temperature dependence of the magnetization shows a sharp drop at $35 K$ in the case of ZFC compounds. For FC materials , the
magnetization does not show any such sharp drops, approaching the constant value. Therefore, applying a magnetic field during preparation of the  material destroys the partial order transition. This suggests that applying magnetic field during preparation of $La_{1-x}Ca_xMnO_3$ one can destroy the AFM-FM transition.

\newpage

\title{Supplementary material to the manuscript \\
	{ Antiferromagnetism to ferromagnetism transition as a transition to partial ordered spins. Application to $La_{1-x}Ca_xMnO_3$ in the doping range $x\geq 0.50$.}}

\begin{abstract}
	
	The calculations leading to the results in the main paper are presented.
	
\end{abstract}

\maketitle

We consider Heisenberg model of two sublattice spin system with antiferromagnetic nearest neighbors exchange $(J)$ of sublattice A and B spins 
%$({\bf S}^A,{\bf S}^B)$
and two different ferromagnetic exchange constants for sublattice A $(J^A)$ and B $(J^B)$ spins. The Hamiltonian of the system is 
\bea \label{FMAFM1}
H & = & J \sum\limits_{\langle ij \rangle} {{\bf S}^A_{i}
	\cdot {\bf S}^B_{j}} \\
& & - J^A\sum\limits_{\ll ij \gg _A } {{\bf S}^A_{i}
	\cdot {\bf S}^A_{j}}\,-\,J^B\sum\limits_{\ll ij \gg _B } {{\bf S}^B_{i}
	\cdot {\bf S}^B_{j}}\nonumber
\eea 
where $({\bf S}^A,{\bf S}^B)$ are spin $s=3/2$ operators, the sums are over all sites of a body-centered cubic (bcc) lattice. To be specific we choose $(J^A>J^B)$. We introduce Holstein-Primakoff representation for the spin operators ${\bf S}^A_{j}(a^+,a)$,
\bea\label{FMAFM2} & &
S_{j}^{A+} = S^{A1}_{j} + i S^{A2}_{j}=\sqrt {2s-a^+_ja_j}\,\,\,\,a_j \nonumber \\
& & S_{j}^{A-} = S^{A1}_{j} - i S^{A2}_{j}=a^+_j\,\,\sqrt {2s-a^+_ja_j}
\\ & & S^{A3}_{j} = s - a^+_ja_j \nonumber \eea
and ${\bf S}^B_{j}(b^+,b)$,
\bea\label{FMAFM3} & &
S_{j}^{B+} = S^{B1}_{j} + i S^{B2}_{j}=-b^+_j\,\,\sqrt {2s-b^+_jb_j}\nonumber \\
& & S_{j}^{B-} = S^{B1}_{j} - i S^{B2}_{j}=-\sqrt {2s-b^+_jb_j}\,\,\,\,b_j
\\ & & S^{B3}_{j} = -s + b^+_jb_j \nonumber. \eea
The operators $a^+_j,\,a_j$ and  $b^+_j,\,b_j$ satisfy the Bose commutation relations.
In terms of the Bose operators and keeping only the quadratic and quartic terms, the effective Hamiltonian
Eq.(\ref{FMAFM1}) adopts the form
\be\label{FMAFM4}H=H_2+H_4\ee where
\bea\label{FMAFM5}
H_2 & = & s J^A\sum\limits_{\ll ij \gg _A }\left( a^+_i a_i\,+\,a^+_j a_j\,-\,a^+_j a_i\,-\,a^+_i a_j\right) \nonumber \\
& + & s J^B\sum\limits_{\ll ij \gg _B }\left( b^+_i b_i\,+\,b^+_j b_j\,-\,b^+_j b_i\,-\,b^+_i b_j\right) \\
& + &s J \sum\limits_{\langle ij \rangle}\left[ b^+_j b_j +  a^+_i a_i -
\left( a^+_i b^+_j+a_i b_j \right)\right] \nonumber
\eea
and
\bea\label{FMAFM6}
H_4 & = & \frac 14 J^A \sum\limits_{\ll ij \gg _A }\left[a^+_i a^+_j( a_i-a_j)^2 + (a^+_i- a^+_j)^2  a_i a_j\right] \nonumber \\
& + & \frac 14 J^B \sum\limits_{\ll ij \gg _B }\left[b^+_i b^+_j( b_i-b_j)^2 + (b^+_i- b^+_j)^2  b_i b_j\right]\nonumber \\
& + & \frac 14 J \sum\limits_{\langle ij \rangle}\left[
a_i b^+_j b_j b_j+a^+_i b^+_j b^+_j b_j  \right. \\
& + & \left. a^+_i a_i a_i b_j+a^+_i a^+_i a_i b^+_j  - 4 a^+_i a_i b^+_j b_j \right]. \nonumber
\eea
The terms without operators are dropped.

The next step is to represent the Hamiltonian in the Hartree-Fock  approximation
\be
H \approx H_{HF}=H_{cl}+H_q, \label {FMAFM7} \ee
with
\bea\label{FMAFM8} H_{cl} & = & 6 N J^A s^2 (u^A-1)^2+ 6 N J^B s^2 (u^B-1)^2 \nonumber \\
& + & 8 N J s^2 (u-1)^2,
\eea
and 
\begin{equation}\label{FMAFM9}
H_q = \sum\limits_{k\in B_r}\left [\varepsilon^a_k\,a_k^+a_k\,+\,\varepsilon^b_k\,b_k^+b_k\,-
\,\gamma_k \left (a_k^+b_k^+ + b_k a_k \right )\,\right ],
\end{equation}
where $N=N^A=N^B$ is the number of sites on a sublattice. The two equivalent sublattices A and B of the bcc lattice are simple cubic lattices. The wave vector $k$ runs over the reduced  first Brillouin zone $B_r$ of a bcc lattice which is the first Brillouin zone of a simple cubic lattice. 

The dispersions are given by equalities
\begin{eqnarray}\label{FMAFM10}
\varepsilon^a_k & = & 4s J^A u^A \left(3-\cos k_x-\cos k_y - \cos k_z\right)
\,+\,8s\,J u \nonumber\\
\varepsilon^b_k & = & 4s J^B u^B \left(3-\cos k_x-\cos k_y - \cos k_z\right)
\,+\,8s\,J u \nonumber\\
\gamma_k & = & 8J\,u\,s\,\cos \frac {k_x}{2} \,\cos \frac {k_y}{2} \,\cos \frac {k_z}{2} \end{eqnarray}
The equations (\ref{FMAFM10}) show that Hartree-Fock parameters ($u^A,u^B,u$) renormalize the intra and inter-sublattice exchange constants
($J^A, J^B,J$) respectively.

To diagonalize the Hamiltonian one introduces new Bose fields
$\alpha_k,\,\alpha_k^+,\,\beta_k,\,\beta_k^+$ by means of the
transformation
\begin{eqnarray}\label{fc-ferri6} & &
a_k\,=u_k\,\alpha_k\,+\,v_k\,\beta^+_k\qquad
a_k^+\,=u_k\,\alpha_k^+\,+\,v_k\,\beta_k
\nonumber \\
\\
& & b_k\,=\,u_k\,\beta_k\,+\,v_k\,\alpha^+_k\qquad
b_k^+\,=\,u_k\,\beta_k^+\,+\,v_k\,\alpha_k, \nonumber
\end{eqnarray}
where the coefficients of the transformation $u_k$ and $v_k$ are real functions of the wave vector $k$:
\bea \label{ferri11} & &
u_k\,=\,\sqrt{\frac 12\,\left (\frac
	{\varepsilon^a_k+\varepsilon^b_k}{\sqrt{(\varepsilon^a_k+\varepsilon^b_k)^2-4\gamma^2_k}}\,+\,1\right
	)}\nonumber \\
\\
& & v_k\,=\,sign (\gamma_k)\,\sqrt{\frac 12\,\left (\frac
	{\varepsilon^a_k+\varepsilon^b_k}{\sqrt{(\varepsilon^a_k+\varepsilon^b_k)^2-4\gamma^2_k}}\,-\,1\right
	)}.\nonumber \eea
The transformed Hamiltonian adopts the form
\begin{equation}
\label{FMAFM2}
H_q = \sum\limits_{k\in B_r}\left(
E^{\alpha}_k\,\alpha_k^+\alpha_k\,+\,E^{\beta}_k\,\beta_k^+\beta_k\,+\,E^0_k\right),
\end{equation}
with new dispersions
\begin{eqnarray}  \label{fc-ferri8} 
& & E^{\alpha}_k\,=\,\frac12\, \sum\limits_{k\in B_r}\left
[\sqrt{(\varepsilon^a_k\,+\,\varepsilon^b_k)^2\,-\,4\gamma^2_k}\,-\,\varepsilon^b_k\,+\,\varepsilon^a_k\right] \nonumber \\
& & E^{\beta}_k\,=\,\frac
12\, \sum\limits_{k\in B_r}\left [
\sqrt{(\varepsilon^a_k\,+\,\varepsilon^b_k)^2\,-\,4\gamma^2_k}\,+\,\varepsilon^b_k\,-\,\varepsilon^a_k\right]
\nonumber \end{eqnarray}
and vacuum energy
\begin{equation}\label{fc-ferri9}
E^{0}_k\,=\,\frac
12\,\left [
\sqrt{(\varepsilon^a_k\,+\,\varepsilon^b_k)^2\,-\,4\gamma^2_k}\,-\,\varepsilon^b_k\,-\,\varepsilon^a_k\right]
\end{equation}
For positive values of the Hartree-Fock parameters and all values of $k\in B_r$,\,the dispersions are nonnegative $ E^{\alpha}_k\geq 0,\, E^{\beta}_k \geq 0$. 
The $\alpha_k$ and $\beta_k$ bosons are  the long-range \textbf{(magnon)} excitations with linear dispersion $E^{\alpha}_k\propto c^{\alpha} |k|$ and $E^{\beta}_k\propto c^{\beta} |k|$, near the zero wave vector. For spin velocity constants we obtained
\begin{equation}\label{FMAFM3}
c^{\alpha} = c^{\beta} = 4s\sqrt{Ju\left(J^A u^A + J^B u^B + J u\right)}.
\end{equation}

The free energy of a system with Hamiltonian $H_{HF}$ equations (\ref{FMAFM7}), (\ref{FMAFM8}) and  (\ref{FMAFM9}) is
\begin{eqnarray}\label{fc-ferri10}
\mathcal{F} & = & 6 N J^A s^2 (u^A-1)^2+ 6 N J^B s^2 (u^B-1)^2 \nonumber \\
& + & 8 N J s^2 (u-1)^2 + \frac 1N \sum\limits_{k\in B_r}E^{0}_k \\
& + & \frac {1}{\beta N} \sum\limits_{k\in B_r}\left[ \ln\left(1-e^{-\beta E^{\alpha}_k}\right)\,+\,\ln\left(1-e^{-\beta E^{\beta}_k}\right)\right],\nonumber\end{eqnarray}
where $\beta\,=\,1/T$\,\, is the inverse temperature.
Then, the system of equations for the Hartree-Fock parameters is
\begin{equation}\label{fc-ferri11}\partial\mathcal{F}/\partial u^A=0,\quad \partial\mathcal{F}/\partial u^B=0,\quad\partial\mathcal{F}/\partial u=0.\end{equation}
One can write them by means of the Bose functions $n_k^{\alpha}$ and $n_k^{\beta}$ of $\alpha$ and $\beta$ excitations.
\bea\label{fc-ferri11a} u^A & = & 1-\frac {1}{3s} \frac 1N \sum\limits_{k\in B_r} \varepsilon_k \left[u_k^2 \,n_k^{\alpha}\, +\, v_k^2\, n_k^{\beta}\, +\, v_k^2\right]\nonumber \\
u^B & = & 1-\frac {1}{3s} \frac 1N \sum\limits_{k\in B_r} \varepsilon_k \left[v_k^2 \,n_k^{\alpha}\, +\, u_k^2\, n_k^{\beta}\, +\, v_k^2\right]\nonumber \\
u & = & 1-\frac 1N \sum\limits_{k\in B_r} \left[\frac {1}{2s}\left(u_k^2 \,n_k^{\alpha}\, +\, v_k^2\, n_k^{\beta}\, +\, v_k^2\right)\right. \\
& + & \left. \frac {1}{2s} \left(v_k^2 \,n_k^{\alpha}\, +\, u_k^2\, n_k^{\beta}\, +\, v_k^2\right)\right. \nonumber \\
& - & \left.8 J u\left(1+n_k^{\alpha}+n_k^{\beta}\right) \frac {\left(\cos \frac {k_x}{2} \cos \frac {k_y}{2} \cos \frac {k_z}{2} \right)^2}{\sqrt{(\varepsilon^a_k\,+\,\varepsilon^b_k)^2\,-\,4\gamma^2_k}}
\right]\nonumber
\eea

The Hartree-Fock parameters are positive functions of $T/J$ , solution of the system of equations (\ref{fc-ferri11}) (see Eqs.(\ref{fc-ferri11a})). Utilizing these functions, one can calculate the spontaneous magnetization on the two sublattices
\begin{eqnarray} \label{fc-ferri12}
M^A & = & <S^{A3}_{j}> \,\,\, j\,\, is\,\, from\,\, sublattice\,\, A \nonumber \\
\\
M^B & = & <S^{B3}_{j}> \,\,\, j\,\, is\,\, from\,\, sublattice\,\, B \nonumber \end{eqnarray}
and $M\,=\,M^A\,+\,M^B$, the spontaneous magnetization of the system.
In terms of the Bose functions 
\begin{eqnarray}\label{fc-ferri13}
M^A & = & s\,-\,\frac 1N \sum\limits_{k\in B_r} \left[u_k^2 \,n_k^{\alpha}\, +\, v_k^2\, n_k^{\beta}\, +\, v_k^2\right] \\
M^B & = & - \,s\,+\,\frac 1N \sum\limits_{k\in B_r} \left[v_k^2 \,n_k^{\alpha}\, +\, u_k^2\, n_k^{\beta}\, +\, v_k^2\right],\nonumber \end{eqnarray}
where $u_k$ and $v_k$ are functions of the wavevector $k$, coefficients in the transformation (\ref{fc-ferri6}).

The Hartree-Fock parameters as fuctions of $T/J$ within interval $(0,T_N/J)$ are depicted in Fig.(1). 
\begin{figure}[!ht]
	\centering\includegraphics[width=4in]{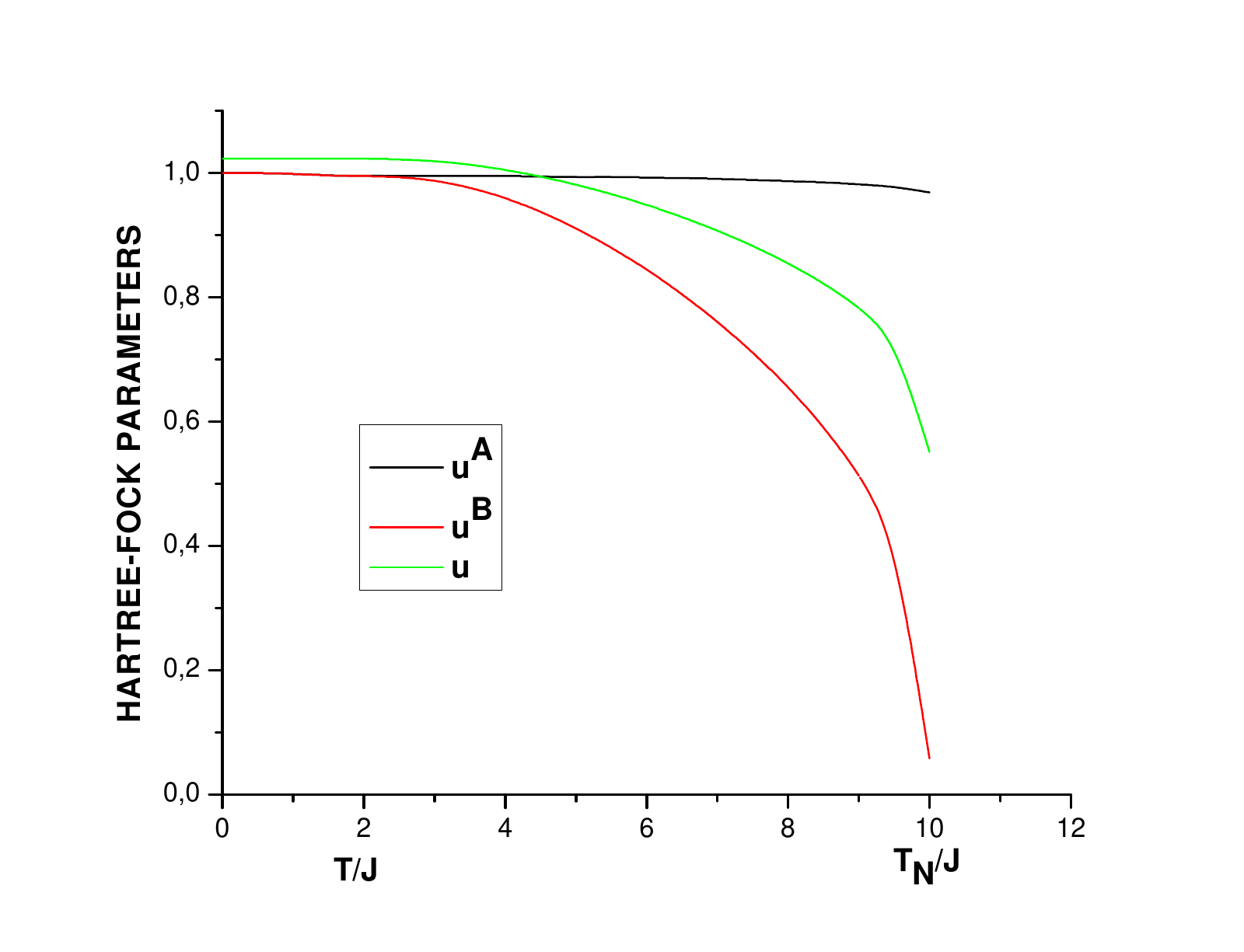}
	%	\epsfxsize=\linewidth
	%	\epsfbox{(fig1)SuppAFMtoFM-HF}
	%	\vskip -6 cm
	\caption{The Hartree-Fock parameters are depicted, as a function of dimensionless temperature $T/J$ within the interval $(0,T_N/J)$, for a system with parameters $J^A/J=0.8$ and  $J^B/J=0.006$.}\label{fig1-HF}       
\end{figure}

The sublattice  A and B magnetization $M^A$,  $M^B$ are depicted in Fig.(2). 
\begin{figure}[!ht]
	\centering\includegraphics[width=4in]{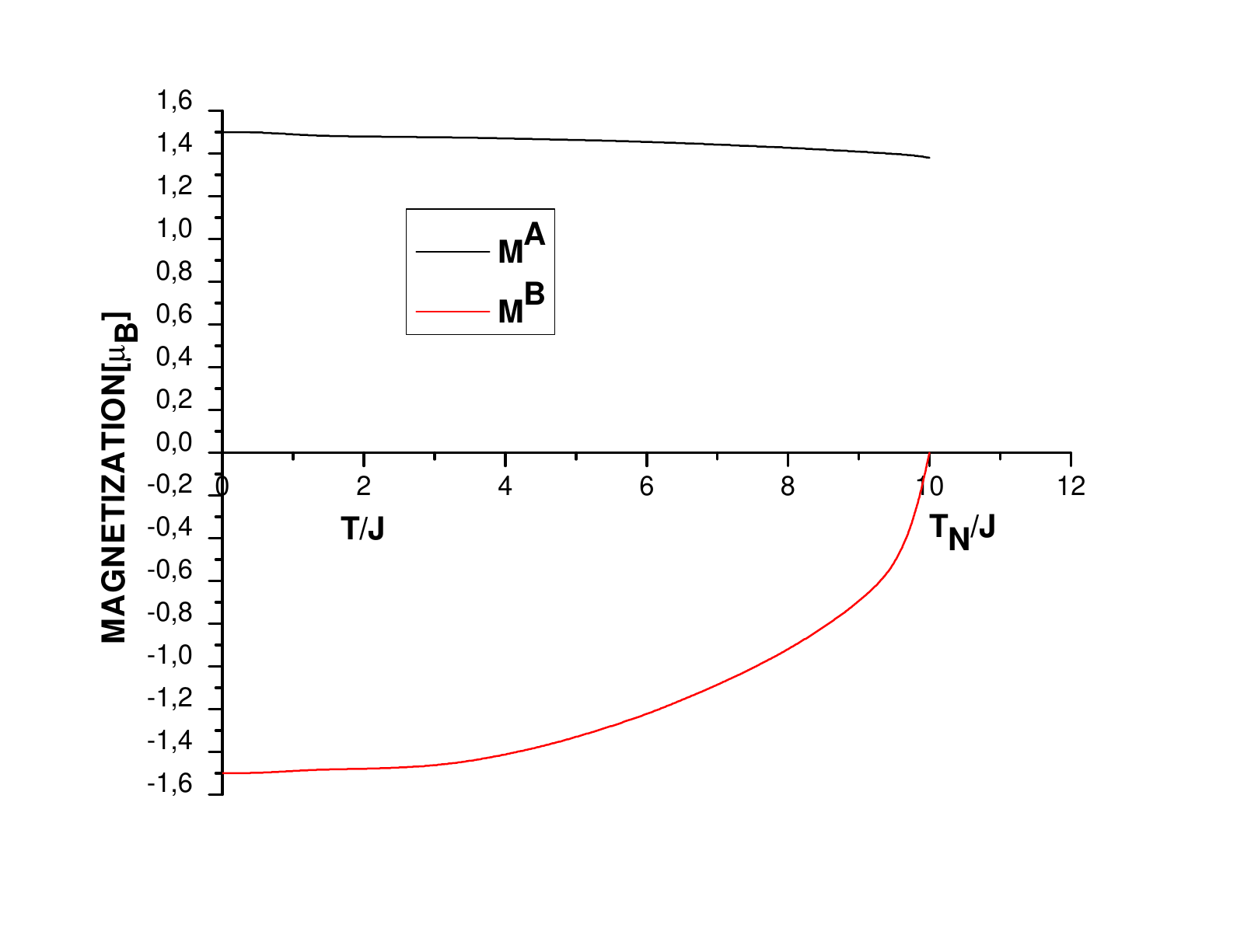}
	%	\epsfxsize=\linewidth
	%	\epsfbox{(fig2)SuppAFMtoFM-MM}
	%	\vskip -6 cm
	\caption{The sublattice $A$ and $B$ magnetization are depicted as a function of dimensionless temperature $T/J$ within the interval $(0,T_N/J)$, for a system with parameters $J^A/J=0.8$ and  $J^B/J=0.006$.}\label{fig2-MM}       
\end{figure}
The magnons' fluctuations suppress in a different way the magnetization on sublattices $A$ and $B$. Quantitatively this depends on the exchange constants $J^A$ and $J^B$ . At characteristic temperature $T_N$  spontaneous magnetization on sublattice $B$ becomes equal to zero, while spontaneous magnetization on sublattice $A$ is still nonzero Fig.(2). The total magnetization $M^A+M^B$ increases and reaches its maximum value at $T_N$ Fig.(2). The system in the temperature interval $(0,T_N)$ has two antiferromagnetic magnons but sublattice A and B magnetization do not compensate each other. We name this phase noncompensated antiferromagnetic. 

Above the N\'eel temperature the magnetic moments of the sublattice B electrons do not contribute the magnetization of the system. To study this "partial ordered" state we first consider the paramagnetic phase.To this end we make use of the Takahashi modified spin-wave theory \cite{Takahashi87,Karchev08}  and introduce two
parameters $\lambda^A$ and $\lambda^B$ to enforce the magnetization on the two sublattices to be equal to zero. The new Hamiltonian is obtained from the old one  (\ref{FMAFM1}) adding two new terms:
\begin{equation}
\label{fc-ferri14} \hat{H}\,=\,H\,-\,\sum\limits_{i\in A}
\lambda^A S^{A3}_{i}\,+\,\sum\limits_{i\in B} \lambda^B S^{B3}_{i}. \end{equation}
In momentum space
the new Hamiltonian adopts the form \be \label{ferri27}
\hat{H} = \sum\limits_{k\in B_r}\left [\hat{\varepsilon}^a_k\,a_k^+a_k\,+\,\hat{\varepsilon}^b_k\,b_k^+b_k\,-
\,\gamma_k\,(b_k a_k+b_k^+ a_k^+)\right] \ee
where the new dispersions are
\be \label{ferri28}
\hat{\varepsilon}^a_k\,=\varepsilon^a_k\,+\,\lambda^A, \qquad
\hat{\varepsilon}^b_k\,=\varepsilon^b_k\,+\,\lambda^B.\ee
It is convenient to represent the parameters
$\lambda^A$ and $\lambda^B$ in the form
\begin{equation} \label{fc-ferri15}
\lambda^A\,=\,8 J u s (\mu^A\,-\,1),\quad
\lambda^B\,=\,8 J u s (\mu^B\,-\,1). \end{equation}
The dispersions $\hat{\varepsilon}^a_k$ and $\hat{\varepsilon}^b_k$ adopt the form 
\begin{eqnarray}\label{FMAFM20}
\hat{\varepsilon}^a_k & = & 4s J^A u^A \left(3-\cos k_x-\cos k_y - \cos k_z\right)
\,+\,8s\,J u \mu^A \nonumber\\
\hat{\varepsilon}^b_k & = & 4s J^B u^B \left(3-\cos k_x-\cos k_y - \cos k_z\right)
\,+\,8s\,J u \mu^B 
\end{eqnarray}
They are positive ($\hat{\varepsilon}^a_k>0$, $\hat{\varepsilon}^b_k>0$) for all values of the wavevector $k$, if the parameters  $\mu^A$ and $\mu^B$ are positive ($\mu^A>0,\,\mu^B>0$).

To obtain the Hamiltonian (\ref{ferri27}) in diagonal form, we use a transformation (\ref{fc-ferri6},\ref{ferri11})  
replacing $\varepsilon^a_k$ and  $\varepsilon^b_k$ with  $\hat{\varepsilon}^a_k$ and $\hat{\varepsilon}^b_k$. The result is
\be
\label{ferri30} \hat{H} = \sum\limits_{k\in B_r}\left
(\hat{E}^{\alpha}_k\,\alpha_k^+\alpha_k\,+\,\hat{E}^{\beta}_k\,\beta_k^+\beta_k+\hat{E}^0_k\right),
\ee 
where
\bea\label{ferri31} & & \hat{E}^{\alpha}_k\,=\,\frac
12\,\left [
\sqrt{(\hat{\varepsilon}^a_k\,+\,\hat{\varepsilon}^b_k)^2\,-\,4\gamma^2_k}\,-\,\hat{\varepsilon}^b_k\,+\,\hat{\varepsilon}^a_k\right] \nonumber \\
& & \hat{E}^{\beta}_k\,=\,\frac
12\,\left [
\sqrt{(\hat{\varepsilon}^a_k\,+\,\hat{\varepsilon}^b_k)^2\,-\,4\gamma^2_k}\,+\,\hat{\varepsilon}^b_k\,-\,\hat{\varepsilon}^a_k\right]\\
& & \hat{E}^{0}_k\,=\,\frac
12\,\left [
\sqrt{(\hat{\varepsilon}^a_k\,+\,\hat{\varepsilon}^b_k)^2\,-\,4\gamma^2_k}\,-\,\hat{\varepsilon}^b_k\,-\,\hat{\varepsilon}^a_k\right]\nonumber\eea
The dispersions Eq.(\ref{ferri31}) are well defined if square-roots in equations
(\ref{ferri31}) are well defined. This is true if 
\be\label{ferri34} \mu^A\mu^B\geq1.\ee
If $\mu^A\mu^B >1$ the dispersions (\ref{ferri31}) are positive ($\hat{E}^{\alpha}_k>0, \hat{E}^{\beta}_k>0$). This means that the system is in paramagnetic phase. 
When  $\mu^A\mu^B =1$ the spectrum of the system posses long-range (magnon) excitation and it is in ordered phase. 

In the partial ordered phase above $T_N$ with $M^B=0$ sublattice A magnetization contributes the total magnetic moment. To study this ordered phase one has to solve a system of four equations, three for Hartree-Fock paraneters and one $(M^B=0)$ for $\mu^A=1/\mu^B$. The solution of the system, the Hartree-Fock parameters and parameters $\mu^A=1/\mu^B$ as fuctions of $T/J$ within interval $(0,T_C/J)$ are depicted in Fig.(3) and Fig.(4) respectively. Above the $T_C$, the critical order-disorder transition temperature, one obtains $\mu^A\mu^B>1$. The sublattice A and B magnetization $M^A$ and $M^B$ are depicted in Fig.(5). The total magnetization $M^A+M^B$ is depicted in Fig.(1) in the main article.

The figure (4) shows that $\mu^B$ is larger than $\mu^A$. As a result we obtain that  $\beta$ excitation is gapped ($E^{\beta}_k>0$ for all values of the wave vector $k$), while $E^{\alpha}_0=0$ and near the zero wave vector
\be \label{ferri35b}
\hat{E}^{\alpha}_k\approx \hat{\rho} k^2\ee 
with spin-stiffness constant
\be \label{ferri35c} \hat{\rho}\,=\,s\frac {(\mu^A+\mu^B)(J^A u^A +J^B u^B)+2J U}{ \mu^B-\mu^A} + sJ^A u^A -sJ^B u^B.\ee 
This means that  $\alpha_k$ boson is the long-range excitation (ferromagnetic magnon) in the system.
\begin{figure}[!ht]
	\centering\includegraphics[width=4in]{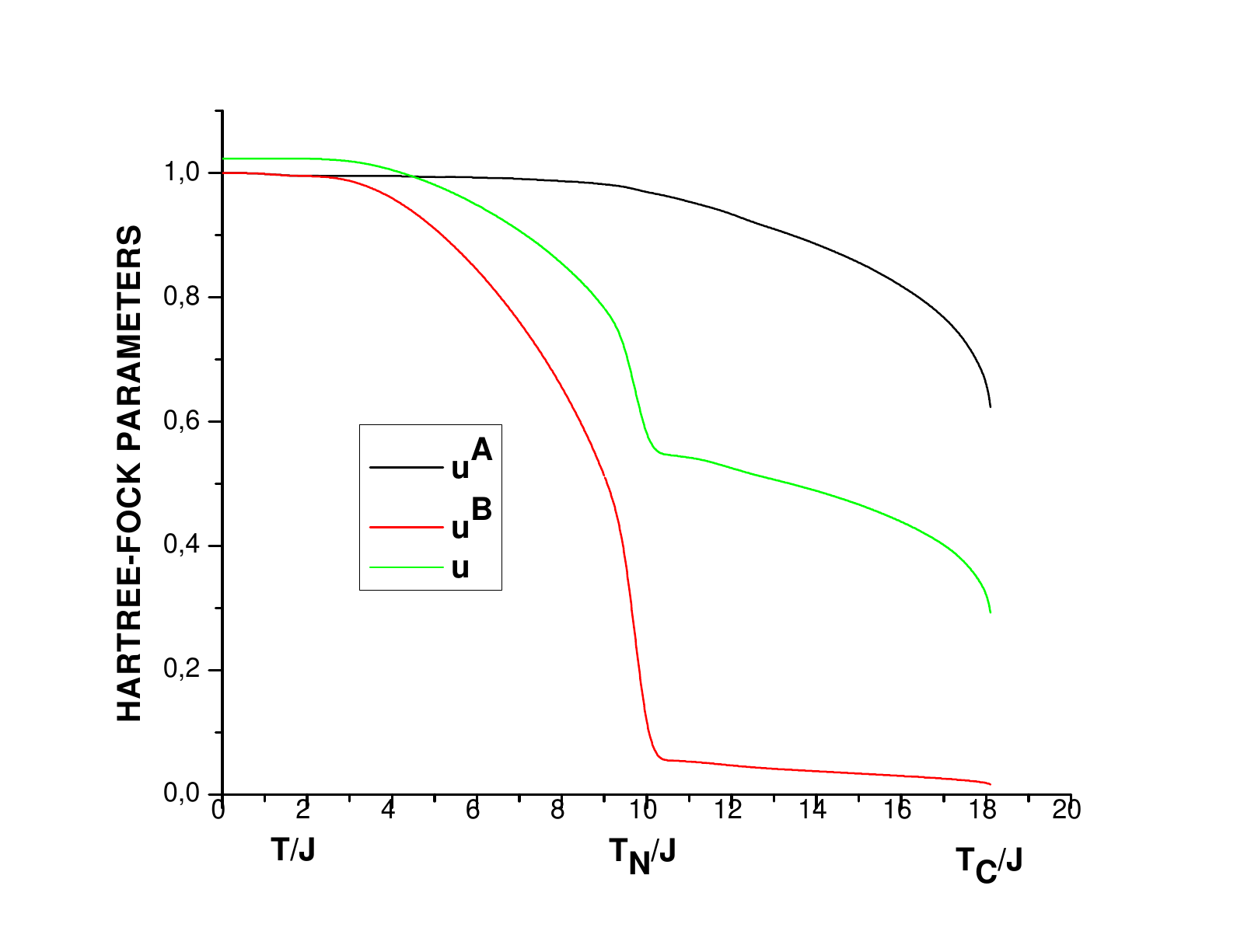}
	%	\epsfxsize=\linewidth
	%	\epsfbox{(fig3)SuppAFMtoFM-HF}
	%	\vskip -6 cm
	\caption{The Hartree-Fock parameters are depicted as a function of dimensionless temperature $T/J$ within the interval $(0,T_C/J)$, for a system with parameters $J^A/J=0.8$ and  $J^B/J=0.006$.}\label{fig3-HF}        
\end{figure}

\begin{figure}[!ht]
	\centering\includegraphics[width=4in]{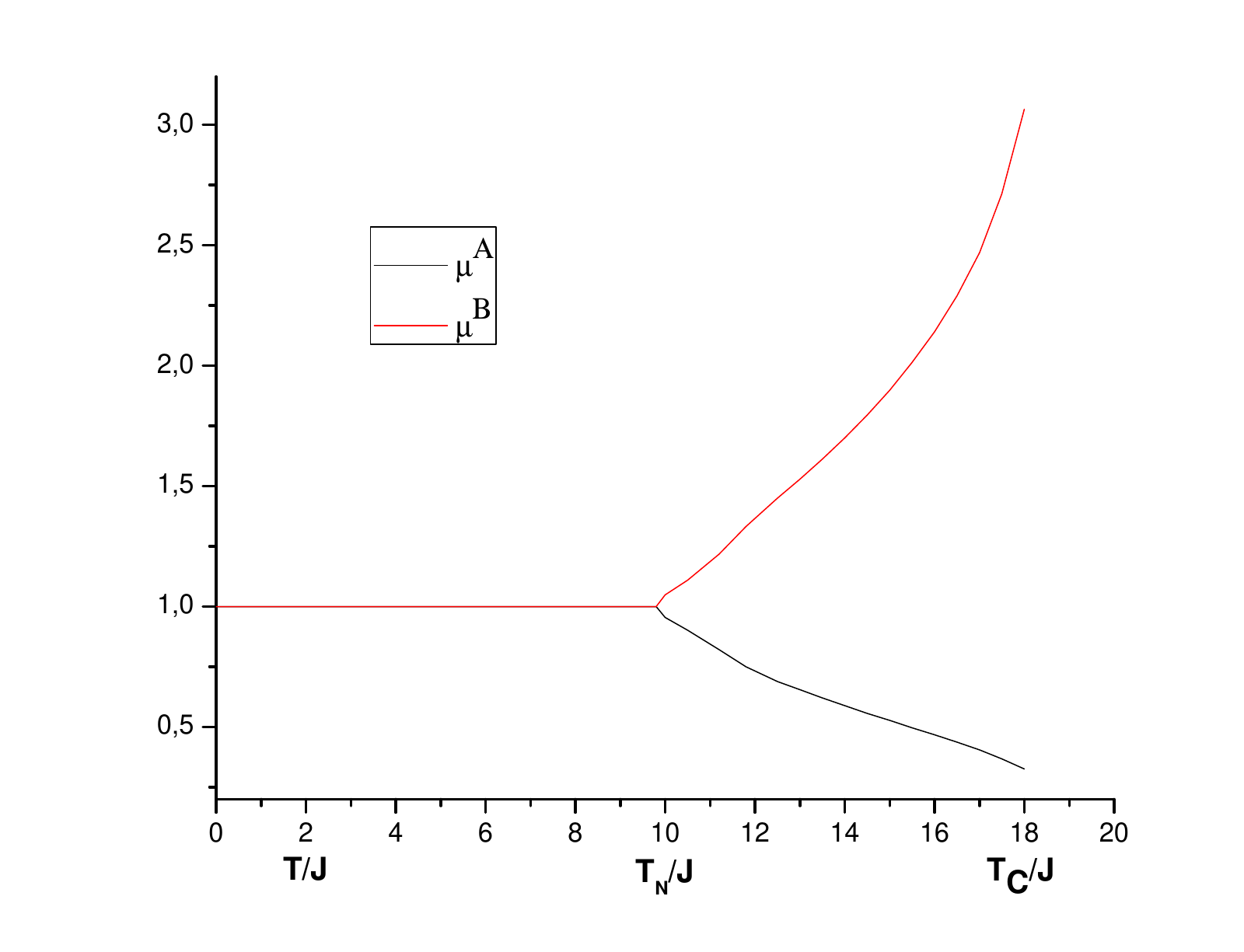}	
	%\epsfxsize=\linewidth
	%\epsfbox{(fig4)SuppAFMtoFM-mu}
	%\vskip -6 cm
	\caption{The parameters $\mu^A$ and $\mu^B$ are depicted as a function of dimensionless temperature 
		$T/J$ within the interval $(0,T_C/J)$, for a system with parameters $J^A/J=0.8$ and  $J^B/J=0.006$.}\label{fig4-mu}       
\end{figure}

\begin{figure}[!ht]
	\centering\includegraphics[width=4in]{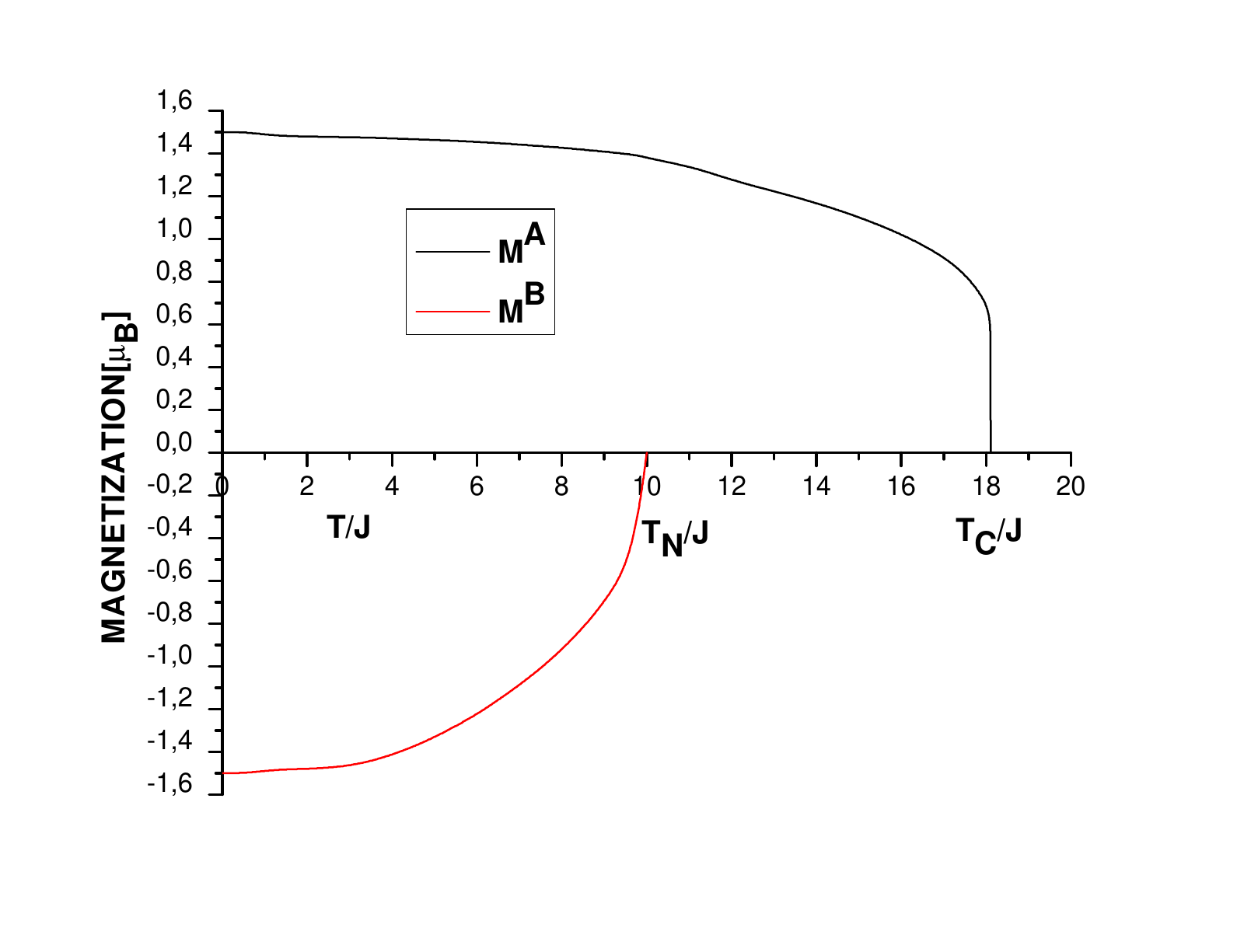}
	%	\epsfxsize=\linewidth
	%	\epsfbox{(fig5)SuppAFMtoFM-MM}
	%	\vskip -6 cm
	\caption{The sublattice $A$ and $B$ magnetization are depicted as a function of dimensionless temperature $T/J$ within the interval $(0,T_C/J)$, for a system with parameters $J^A/J=0.8$ and  $J^B/J=0.006$.}\label{fig5-MM}       
\end{figure}

In conclusion, we will note that in the paramagnetic phase $\mu^A\mu^B>1$. On decreasing the temperature the product $\mu_1\mu_2$ decreases, remaining larger than one. At temperature $T_C$ at which the product becomes equal to one ($\mu^A\mu^B=1$) the system undergoes paramagnetic-ferromagnetic phase transition. Only sublattice A spins contribute the total magnetization, i.e. the ferromagnetic phase is partial ordered phase. Below $T_C$ at $T_N$, $\mu^A$ and $\mu^B$ become equal to one $(\mu^A=\mu^B=1)$ and the system undergoes ferromagnetic to noncompensated  antiferromagnetic phase transition. In the low temperature phase sublattice A and B spins contribute the total magnetization. Hence, ferromagnetic-antiferromagnetic phase transition is a partial ordered transition. 

It is usfull to plot magnon dispersions at different temperature. The dimensionless dispersions $E^{\alpha}/2sJ$ and  $E^{\beta}/2sJ$, below $T_N$ are plotted in figure 6  for system with parameters $J^A/J=0.8$, $J^B/J=0.5$ and $T=0.5 T_N$  The dispersions at temperature above $T_N$, $T=1.5 T_N$ are ploted in figure 7.

\begin{figure}[!ht]
	\centering\includegraphics[width=7in]{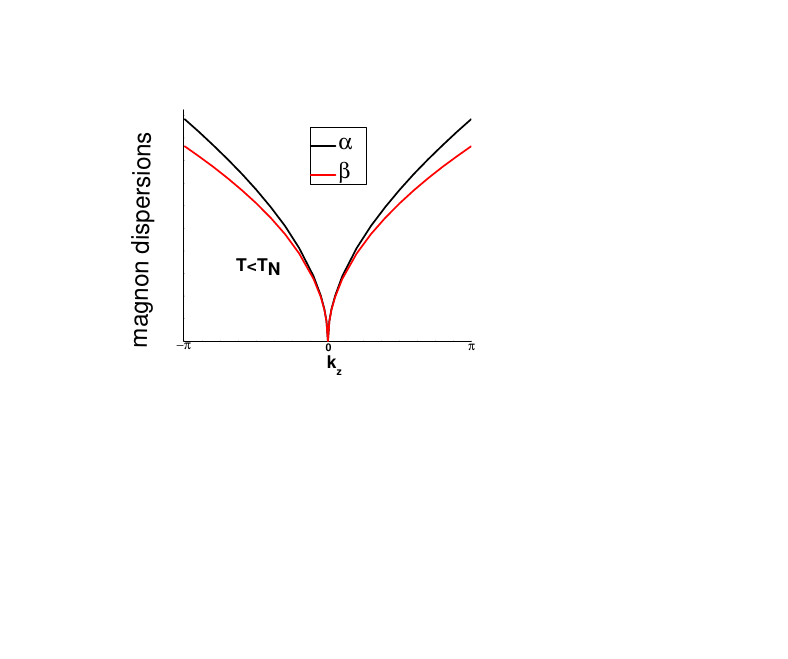}\label{md1} 
	%	\epsfxsize=\linewidth
	%	\epsfbox{(fig5)SuppAFMtoFM-MM}
	\vskip -6 cm
	\caption{The dimensionless dispersion $E^{\alpha}/2sJ$ and $E^{\beta}/2sJ$ are depicted  as a function of $k_z$ and $k_x=k_y=0$, for system with parameters  $J^A/J=0.8$, $J^B/J=0.5$ and $T=0.5 T_N$}     
\end{figure}

\begin{figure}[!ht]
	\centering\includegraphics[width=7in]{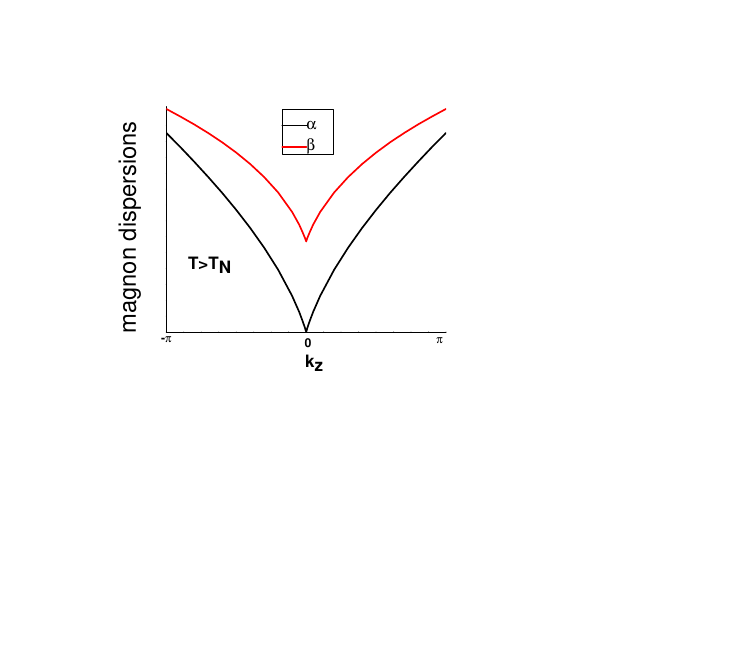}\label{md2} 
	%	\epsfxsize=\linewidth
	%	\epsfbox{(fig5)SuppAFMtoFM-MM}
	\vskip -6 cm
	\caption{The dimensionless dispersion $E^{\alpha}/2sJ$ and $E^{\beta}/2sJ$ are depicted  as a function of $k_z$ and $k_x=k_y=0$, for system with parameters  $J^A/J=0.8$, $J^B/J=0.5$ and $T = 1.25 T_N$}         
\end{figure}

\end{document}